\documentclass[conference]{basi}

\usepackage[T1]{fontenc}
\usepackage[british]{babel}
\usepackage[varg]{txfonts}

\usepackage{rotating}
\usepackage{dcolumn}

\begin{document}
\title[X-ray and IR properties of Be/X-ray pulsars]{X-ray and Infrared Properties of Be/X-ray Binary Pulsars}
\author[S.~Naik]{Sachindra Naik\thanks{email: \texttt{snaik@prl.res.in}}\\
Physical Research Laboratory, Navrangapura, Ahmedabad 380009, Gujarat, India}

\pubyear{2013}
\volume{**}
\pagerange{**--**}

\date{Received --- ; accepted ---}

\maketitle


\begin{abstract}
During normal Type~I outbursts, the pulse profiles of Be/X-ray binary pulsars
are found to be complex in soft X-ray energy ranges. The profiles in soft X-ray
energy ranges are characterized by the presence of narrow absorption dips or 
dip-like features at several pulse phases. However, in hard X-ray energy ranges, 
the pulse profiles are rather smooth and single-peaked. Pulse phase-averaged 
spectroscopy of the these pulsars had been carried out during Type~I outbursts. 
The broad-band spectrum of these pulsars were well described by partial covering 
high energy cutoff power-law model with interstellar absorption and Iron K$_\alpha$ 
emission line at 6.4 keV. Phase-resolved spectroscopy revealed that the presence 
of additional matter at certain pulse phases that partially obscured the emitted 
radiation giving rise to dips in the pulse profiles. The additional absorption is 
understood to be taking place by matter in the accretion streams that are phase 
locked with the neutron star. Optical/infrared observations of the companion Be
star during these Type~I outbursts showed that the increase in the X-ray intensity
of the pulsar is coupled with the decrease in the optical/infrared flux of the
companion star. There are also several changes in the IR/optical emission line profiles 
during these X-ray outbursts. The X-ray properties of these pulsars during Type~I
outbursts and corresponding changes in optical/infrared wavebands are briefly 
discussed in this paper.
\end{abstract}


\section{Introduction}\label{s:intro}

High Mass X-ray Binaries (HMXBs) are known to be strong X-ray emitters and appear
as the brightest X-ray sources in the sky. These systems are classified as Be/X-ray
binaries (largest subclass of HMXBs) and supergiant X-ray binaries. The compact object 
in Be/X-ray binaries is generally a neutron star where as the companion is a B or 
O-type star which shows Balmer emission lines in its spectra. The neutron star in these 
systems is typically in a wide orbit with moderate eccentricity with orbital period in 
the range of 16-400 days. The neutron star spends most of the time far away from the 
circumstellar disk of the companion Be star. It accretes matter from the companion 
while passing through its circumstellar disk at the periastron passage. The abrupt 
accretion of huge amount of matter onto the neutron star results in strong outbursts 
(Okazaki \& Negueruela 2001) during which the X-ray emission from the pulsar can be 
transiently enhanced by a factor more than $\sim$10. Pulsars in Be/X-ray binary systems 
generally show periodic normal (Type~I) X-ray outbursts that coincide with the periastron 
passage of the neutron star and giant (Type~II) X-ray outbursts which do not show any 
clear orbital dependence (Negueruela et al. 1998). The spin period of these pulsars is 
found to be in the range of a few seconds to several hundred seconds. The X-ray spectra 
of these pulsars are generally hard. Fluorescent iron emission line at 6.4 keV is observed 
in the spectrum of most of the accretion powered X-ray pulsars. For a brief review on
the properties of Be/X-ray binary pulsars, refer to Paul \& Naik (2011).

The Be stars in the Be/X-ray binary systems show spectral lines such as hydrogen
(Balmer and Paschen series) lines in emission (Porter \& Rivinius 2003). Apart
from these hydrogen emission lines, these stars occasionally show He and Fe lines
in emission (Hanuschik 1996). These Be stars show an infrared excess i.e. an excess 
amount of IR emission compared to the IR emission from an absorption-line B star of 
same spectral type. The observed IR excess and emission lines in the optical/IR spectra
of Be stars are attributed to the presence of a equatorial circumstellar disk. In Be/X-ray
binary systems, the circumstellar disk of the companion star is being truncated/evacuated
by the neutron star at the periastron passage resulting in X-ray outbursts. During X-ray
outbursts, there are several occasions when extreme changes in emission line profiles and
optical/infrared $J, H, K$ magnitudes of companion Be star have been reported. In the 
following sections, we briefly describe the X-ray properties of the pulsar during type~I 
outbursts and related changes observed in optical/infrared wavebands in Be/X-ray binaries.

\begin{figure}
\centering
\includegraphics[height=4.7in, width=2.1in, angle=-90]{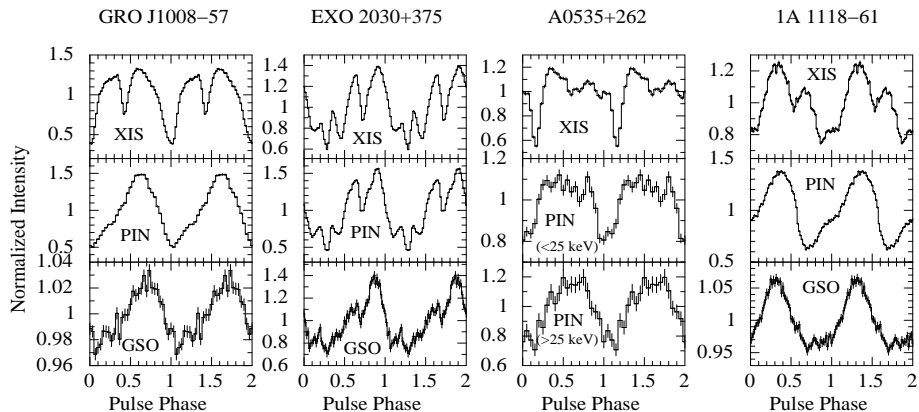}
\caption{The pulse profiles of four Be/X-ray binary pulsars during type~I outbursts.
Profiles of the pulsars obtained from $Suzaku$ XIS, PIN and GSO data are in 0.5-10 keV,
10-70 keV and 40-200 keV energy ranges, respectively. However, for A0535+262, the pulse 
profiles shown in middle and bottom panels are in 10-25 keV and 25-70 keV energy ranges.}
\label{pp}
\end{figure}

\section{Pulse profiles of Be/X-ray pulsars during Type~I outbursts}

\begin{figure}
\centerline{\includegraphics[angle=-90,width=5cm]{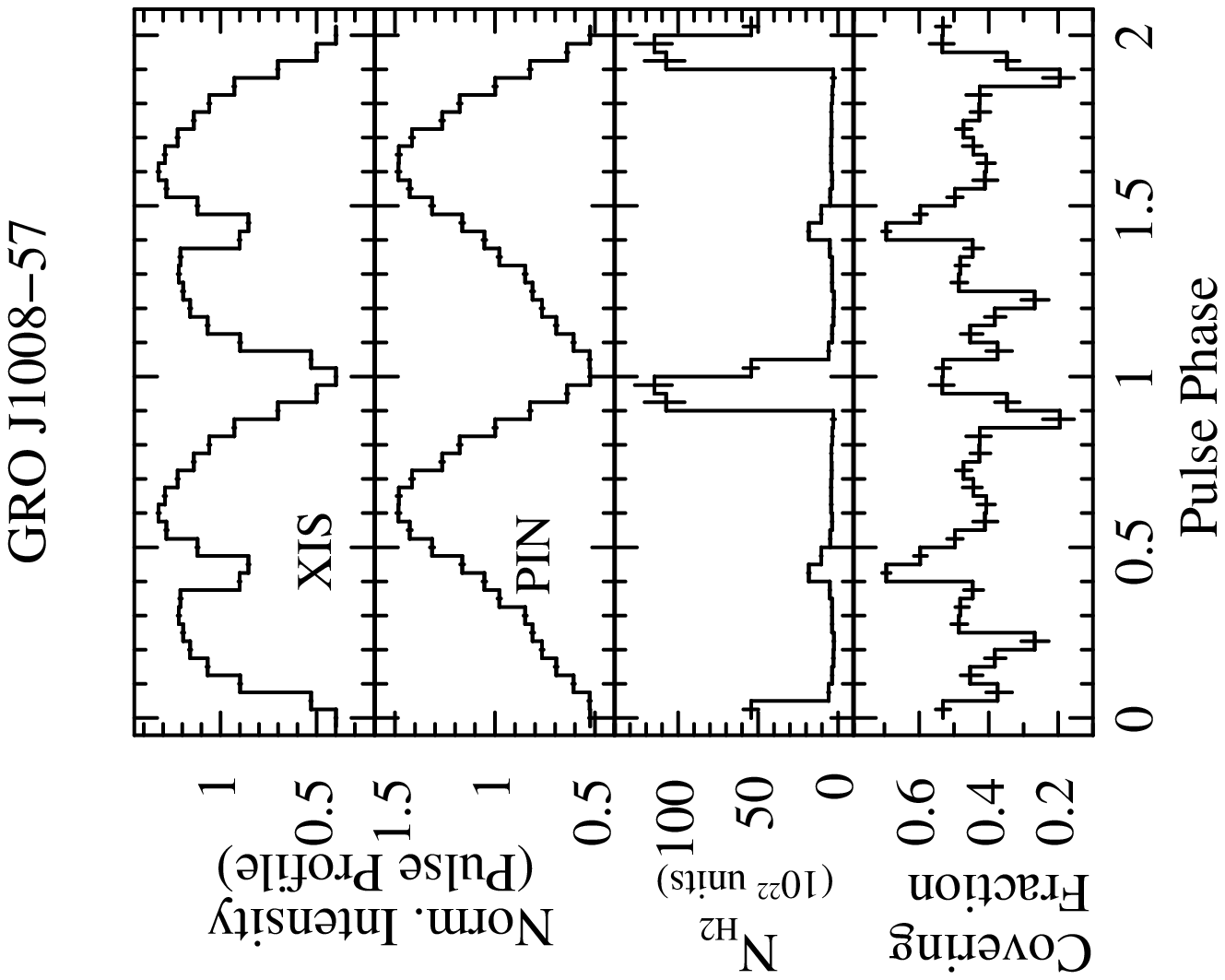}  
            \includegraphics[angle=-90,width=5cm]{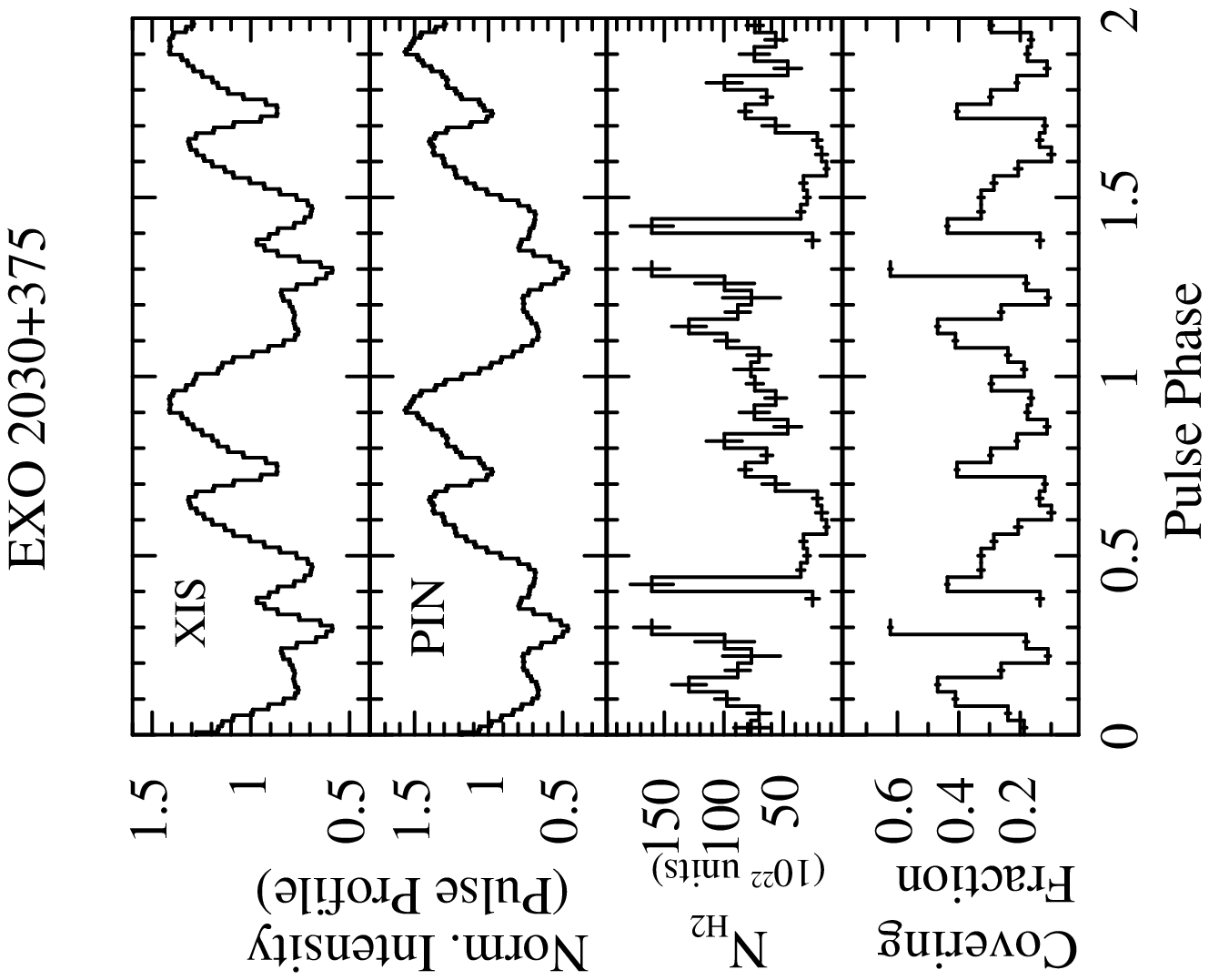}}
\caption{Change in the values of additional absorption column density ($N_{H2}$; in units 
of 10$^{22}$ atoms cm$^{-2}$) (local to the neutron star) and covering fraction with pulse 
phase for Be/X-ray binary pulsars GRO~J1008-57 and EXO~2030+375. Pulse profiles in 0.5-10 
keV (XIS) and 10-70 keV (PIN) energy ranges are shown in top two panels for both the pulsars.}
\label{phrs}
\end{figure}

Accretion powered transient Be/X-ray binary pulsars show luminosity dependence of pulse 
profiles. During quiescent phase when the mass accretion rate is steady and low, these 
pulsars show smooth and single-peaked profiles (viz. A0535+262; Mukherjee \& Paul 2005).
However, during the periastron passage, abrupt accretion of huge amount of mass onto the 
neutron star from the Be circumstellar disk results in significant increase in the X-ray
luminosity. During these outbursts, the pulse profile of these pulsars show the presence 
of dips at certain pulse phases. These dips are found to be prominent at soft X-ray energy
ranges and gradually disappear from the hard X-ray pulse profiles. Pulse profiles of 
four Be/X-ray binary pulsars such as GRO~J1008-57 (93.737 s spin period;  Naik et al. 2011), 
EXO~2030+375 (41.41 s spin period; Naik et al. 2013), A0535+262 (103.375 s spin period; Naik 
et al. 2008) and 1A~1118-61 (407.49 s spin period; Maitra et al. 2012) are shown in 
Figure~\ref{pp}. The observations of these pulsars were carried out during respective 
type~I outbursts using XIS, PIN and GSO detectors of $Suzaku$ observatory. The presence 
of dips in soft X-ray pulse profiles (0.5-10 keV range; XIS) are clearly seen in all cases. 
These dips gradually disappear with energy making the hard X-ray pulse profiles (PIN/GSO) 
smooth and single-peaked. Broad-band energy spectrum of these Be/X-ray binary pulsars
during respective type~I outbursts were described by several continuum models such as
high energy cutoff power-law, negative and positive power-law with exponential cutoff,
a partial covering power-law with high energy cutoff continuum models. However, the 
partial covering power-law with high energy cutoff model is found to be best suitable
model to explain both the phase-averaged and phase-resolved spectra. This model was used
to fit the energy spectrum of many other Be/X-ray binary pulsars which are not described
in this paper. Using this model, the dips or dip-like features in the pulse profile
are explained by the presence of an additional absorption component with high column
density and covering fraction at the same pulse phase. Pulse phase-resolved spectroscopy 
of two Be/X-ray binary pulsars (GRO~J1008-57 and EXO~2030+375) are shown in Figure~\ref{phrs}. 
It can be seen that the values of additional absorption column density $N_{H2}$ and covering 
fraction are high at dip phases in the pulse profiles. The additional absorption is
understood to be taking place by matter in the accretion streams that are phase
locked with the neutron star.

\section{Infrared observations of the Be/X-ray binaries}

During the periastron passage of the neutron star in Be/X-ray binary systems, the 
circumstellar disk of the Be companion star is most affected. The circumstellar
disk is partially truncated of evacuated during the periastron passage. This evacuated
matter from the disk contributes towards the enhancement of the X-ray intensity from
the neutron star giving rise to type~I outbursts. As the circumstellar disk significantly
contributes towards the infrared emission from the companion Be star, the effect of 
periastron passage is expected to be pronounced in IR bands than in optical bands. A 
striking episode of circumstellar disk loss and subsequent formation of new disk have 
been reported in A0535+262/HDE~245770 Be/X-ray binary system (Haigh et al. 1999). During 
this episode, the Br $\gamma$ emission line was detected in absorption along with 
significant decrease in the strength of He~I line at 2.058 $\mu$m. IR spectroscopy of the 
companion star HDE~245770 obtained over 1992-1995 showed significant variability, implying 
changes in the circumstellar disk (Clark et al. 1998). A decrease in the flux of Paschen 
series lines, the strength of $H_\alpha$ line and the optical continuum emission were seen 
between 1993 December and 1994 September. These changes were attributed to the reduction in 
the emission measure of the Be disk. 

\begin{figure}
\centering
\includegraphics[height=3.6in, width=2.4in, angle=-90]{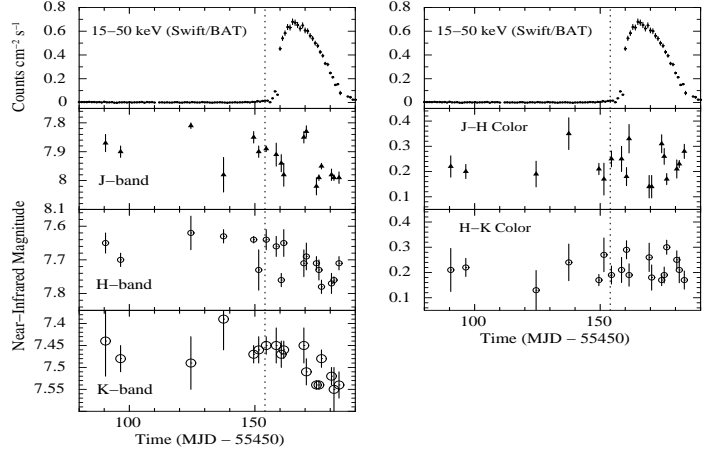}
\caption{The Swift/BAT X-ray light curve (top panels) and the near-infrared $JHK$ 
light curves (left panels) of the Be star in A0535+262/HDE~245770 binary system, 
covering the type~I X-ray outburst in 2011 February-March. The second and third 
panels in right side show the J-H and H-K colors of the Be/X-ray binary system.}
\label{ir}
\end{figure}

Extensive monitoring of this Be/X-ray binary system was carried out in near-infrared 
bands at different phases of its $\sim$111 d orbital period with the 1.2 m telescope 
of Mt. Abu IR observatory (Naik et al. 2012). During this campaign, changes of such 
striking nature i.e. appearance of emission lines in absorption or significant change 
in the emission line intensities were not detected. However, photometric observations 
of the Be star showed a gradual and systematic fading in the $JHK$ light curves since 
the onset of the X-ray outburst in 2011 February-March (Figure~\ref{ir}). 
Interferometric studies of a few Be stars indicate that different emission lines in 
IR and optical spectrum originate from different regions in the circumstellar disk 
(Gies et al. 2007). In our monitoring of A0535+262 Be/X-ray binary system, the strength
of emission lines in $JHK$ spectra were comparable during the X-ray outburst to that
during quiescent phase. However, gradual decrease in the photometric magnitude of the
Be star suggest that a mild evacuation/truncation of the circumstellar disk of the 
companion star took place without affecting the line emitting regions in the disk. 
Near infrared photometric and spectroscopic monitoring of another Be/X-ray binary 
X~Per/HD~24534 in $JHK$ bands revealed that the Be star reached a very high state
of near-IR brightness (Mathew et al. 2012; 2013). The Be star continued brightening 
in $JHK$ bands throughout the monitoring campaign from 2010 December to 2012 April. 
The $JHK$ spectra were dominated by emission lines of He~I and Paschen and Brackett 
lines of H~I. It was found that the equivalent widths and the fluxes of prominent
H~I and He~I emission lines in $JHK$ spectra anti-correlate with the strength of the
continuum. During the monitoring of the Be star from 2010 December to 2012 April, 
however, no X-ray flaring episode was detected. Though the neutron star in the
binary system does not pass through the circumstellar disk of the Be companion
due to low value of eccentricity, there are reports of X-ray outbursts detected
in X~Per from continuous monitoring with $RXTE$, $Swift$ and $INTEGRAL$ observatories 
(Lutovinov et al. 2012). Event of circumstellar disk loss in X~Per has not been
reported since 1989. The observed near-IR brightening of the Be star can be interpreted
as due to increase in the size of the circumstellar disk. In that case, there is
a possibility of detection of an X-ray outburst in X~Per Be binary system in future.  

Extreme changes in emission line profiles of Be stars have been observed in 
several cases. Some notable examples  of drastic H$\alpha$ line profile changes 
are in Omicron Cas (Slettebak \& Reynolds 1978), GX~304-1 (Corbet et al. 1986), 
$\gamma$~Cas (Doazan et al. 1983) etc. In Be/X-ray binary 4U~1258-61 (GX~304-1),
Corbet et al. (1986) found that the H$\alpha$ line profile changed from a shell 
profile to an absorption profile over a period of four years. In V635~Cas Be/X-ray
binary system, the H$\alpha$ line profile was found to change from emission to
absorption during 1997 February - 1997 July (Negueruela et al. 2001). The 
striking feature here is, all these changes are observed to be coincided with 
the type~I X-ray outbursts from the neutron star in the binary systems. 

\section*{Acknowledgments}
The research work at Physical Research Laboratory is funded by the Department
of Space, Government of India. The author thanks the organizers for their invitation 
to participate in the conference. Sincere thanks to Prof. Biswajit Paul of Raman 
Research Institute, Bangalore for kindly agreeing to present this work in 
the conference at very short notice.

\label{lastpage}

\begin{thebibliography}{}
\bibitem{Clark}Clark J. S., Steele I. A., Coe M. J., Roche P., 1998, MNRAS, 297, 657
\bibitem{Corbet}Corbet R. H. D., Smale A. P., Menzies J. W., 1986, MNRAS, 221, 961
\bibitem{Doazan}Doazan V., Franco M., Sedmak G., Stalio R., Rusconi L., 1983, A\&A, 128, 171
\bibitem{Gies}Gies D. R. et al., 2007, ApJ, 654, 527
\bibitem{Haigh}Haigh N. J., Coe M. J., Fabregat J., 1999, MNRAS, 310, L21
\bibitem{Hanuschik}Hanuschik R. W., 1996, A\&A, 308, 170
\bibitem{Lutovinov}Lutovinov A., Tsygankov S., Chernyakova M., 2012, MNRAS, 423, L1978
\bibitem{Maitra}Maitra C., Paul B., Naik S., 2012, MNRAS, 420, 2307
\bibitem{Mathew1}Mathew B., Banerjee D. P. K., Naik S., Ashok N. M., 2012, MNRAS, 423, 2486
\bibitem{Mathew2}Mathew B., Banerjee D. P. K., Naik S., Ashok N. M., 2013, AJ, 145, 158
\bibitem{Muk}Mukherjee U., Paul B., 2005,  A\&A, 431, 667
\bibitem{Naik1}Naik S., et al., 2008, ApJ, 672, 516
\bibitem{Naik2}Naik S., Paul B., Kachhara C., Vadawale S. V., 2011, MNRAS, 413,241
\bibitem{Naik3}Naik S., Mathew B., Banerjee D.~P.~K., Ashok N.~M., Jaiswal R.~R., 
          2012, RAA, 12, 177
\bibitem{Naik4}Naik S., Maitra C., Jaisawal G. K., Paul B., 2013, ApJ, 746, 158
\bibitem{Negueruela}Negueruela I., Reig P., Coe M. J., Fabregat J., 1998, A\&A, 336, 251 
\bibitem{Negueruela}Negueruela I., Okazaki A. T., Fabregat J., Coe M. J., Munari U.,
         Tomov T., 2001, A\&A, 369, 117
\bibitem{Okazaki}Okazaki A. T., Negueruela I., 2001, A\&A, 377, 161
\bibitem{Paul}Paul B., Naik S., 2011, BASI, 39, 429
\bibitem{Porter}Porter J. M., Rivinius T., 2003, PASP, 115, 1153
\bibitem{Slettebak}Slettebak A., Reynolds R. C., 1978, ApJS, 38, 205
\end{thebibliography}
\end{document}